# A photon-like wavepacket with quantised properties based on classical Maxwell's equations

John. E. Carroll *

*Centre for Advanced Photonics & Electronics,*
*9, J.J. Thomson Avenue, University of Cambridge, CB3 0FA , UK*

A photon-like wavepacket based on novel solutions of Maxwell's equations is proposed. It is believed to be the first 'classical' model that contains so many of the accepted quantum features. In this new work, novel solutions to Maxwell's classical equations in dispersive guides are considered where local helical twists with an arbitrary angular frequency Ω modulate a classical mode (angular frequency ω, group velocity $v_g$). The modal field patterns are unchanged, apart from the twist, provided that the helical velocity $v_h$ equals $v_g$. Pairs of resonating retarded and advanced waves with modal and helical frequencies (ω,Ω) and (ω,−Ω) respectively, trap one temporal period of the underlying classical mode forming a photon-like packet provided Ω = (M+1/2)ω : 'Schrödinger' frequencies. This theory supports experimental evidence that the photon velocity does not change with M in dispersive systems. Promotion and demotion increase or decrease the helical frequencies in units of ω. An energy of interaction between retarded and advanced waves in the wave-packet is also proportional to these helical frequencies Ω =(M+1/2)ω similar to Planck's law. Group velocity and polarisation are unaffected by the value of M. Advanced waves enable phase and polarisation to be predicted along all future paths and may help to explain the outcomes of experiments on delayed-choice interference and entanglement, without causality being violated.

## 1. Introduction

Understanding how to engineer photonic secure communications and also quantum computers (Imre and Balazs 2004, Pavicic 2005) provides a strong motivation to find a more intuitive picture that will link the physics of Maxwell's classical equations to the key quantum phenomena of photons. The standard method of quantizing Maxwell's equations associates harmonic oscillators for each modal frequency ω with its propagation vector $\boldsymbol{k}$ (Marcuse 1980, Cohen-tannoudji *et al.* 1989, Loudon 2000, Gerry & Knight 2005). Although Schrödinger frequencies (M+½)ω (M integer) are associated with each harmonic oscillator there is no *physical* explanation of their significance. One might think that the value of M should change the propagation velocity in a dispersive media. Steinberg et al.

---

* electronic address:jec1000@cam.ac.uk

1992 examined centimetric lengths of bulk glass while more recently the author suggested experiments on dispersive fibres (Ingham et al. 2006). There is no experimental evidence that suggests any change in group velocity as one changes from a classical composition of photon states (M >>1) to single photons (M = 1). There is also the difficulty of explaining the counter-intuitive experiments such as delayed-choice interference and entanglement (Hellmuth *et al.* 1987 , Shih and Alley 1988) showing that the outcome of interference and polarisation of a single photon can be predicted long before the energy in the photon arrives. While non-local properties of photons are reinforced by the emphasis on *k*-space in standard quantisation methods, experiments such as those by Hong and Mandel (1986) or Branning et al. (2000) show that the photon can be localised in space or time. The conceptual difficulties of wave-particle duality have been well reviewed by Roychoudhuri and Roy (2003) in a special issue of Optics and Physics News discussing "What is a photon?". A wealth of references can be found to support a wide range of attacks on modelling the photon, its structure, dynamics and modelling [e.g. Cook 1982, Gersten 1999, Hunter & Wadlinger 1989, Jennison 1999, Kobe 1999, Ligare & Oliveri 2002, Stumpf & Borne 2001]. However the photon's conceptual difficulties still persist.

This new theory is believed to be the first model, based on the solid foundations of Maxwell's equations, that offers such close parallels between classical and quantum theory. Even if the reader accepts it only as an analogy, it illuminates many of the difficulties of wave-particle duality. The theory is based on novel classical solutions of Maxwell's equations that seem to have been overlooked. The first sections establish the physics of these solutions dealing with a mode at an angular frequency $\omega$ that is subjected to local helical rotations of all transverse vectors at an angular frequency $\Omega$ that may be much larger and unrelated to $\omega$. Conventional modulation would lead to significant dispersion but this new unconventional method shows that there is no additional dispersion.

The theory starts by using the standard technique of splitting Maxwell's classical equations into TE and TM modes (section 2). A relativistic format is also used in order to demonstrate a new interpretation of the solutions. In section 3 one selects an arbitrary modal solution with a frequency $\omega$ then modulates this mode with an arbitrary local helical rotation of frequency $\Omega$, initially unrelated to $\omega$. It is found that, provided that the helical velocity $v_h$ is equal to the modal group velocity $v_g$, the mode with its frequency $\omega$ and modal propagation constant $k$ can be retained though all the field patterns are now modulated by the helical rotation at angular frequency $\Omega$ about the fibre's axis. This type of local helical twist should not be confused with the orbital helical rotation of the single frequency Laguerre Gaussian modes [Allen et al 1992, Berry 2004, Leach et al 2004 ]. This section 3 also shows that, in spite of the local helical modulation, the *classically calculated* energy of the mode does not change.

Section 4 then shows how to create a new type of wave-packet that traps one whole temporal period of an underlying mode at frequency $\omega$ by having a pair of contra-rotating helical waves with eigen helical frequencies $\Omega = (M+½)\,\omega$ ; M is integer. In this case the polarisation states remain essentially unaltered. The resulting wave-packet travels at the modal group velocity $v_g$ that is independent of M . Section 5 uses the relativistic format of section 2 to explain how two distinct interpretations of Maxwell's equations affect this wave-packet. This discussion revisits advanced and retarded waves, related to seminal discussions by Wheeler and Feynman (1945) and related to left-handed waves discussed by Vesalago (1968). One of the helically modulated waves is associated with a standard retarded wave but the counter rotating helical modulation is postulated to be an advanced wave that by definition in this work carries power from the future towards the past. These two waves overlap in space-time and create a resonant wave-packet where the advanced waves provides the feedback required for a resonance. Advanced waves, as used here, never carry energy by themselves and never violate causality. These new wave-packets are called PRAHM modes (Packets of Retarded and Advanced Helically Modulated modes ). They have eigen frequencies $(M+½)\omega$ and trap one whole period of the underlying classical mode. Section 6 shows that, over the region where the advanced and retarded waves overlap, there is an energy that can be associated with the counter rotating helical modulation. Using Maxwell's classical equations this energy is proportional to $(M+½)\omega$ creating a form of a classical Planck's law. A heuristic estimate can be made of Planck's constant within the right order of magnitude.

Section 7 demonstrates some parallels between this classical Maxwellian theory and standard quantum theory. It is shown for example that promotion, demotion and annihilation all follow as for standard quantum theory but with classical descriptions attached to these processes. The final section, section 8, of the paper rounds up the work with brief discussions on probability, uncertainty, non-local effects and conclusions.

## 2. Maxwell's equations in a circular waveguide

This section has two objectives. The first is to rehearse a well known result, but in the format required here, that Maxwell's equations can be split into Transverse Electric (TE) and Transverse Magnetic (TM) modes as for example in Sander and Reed (1986) [see appendix A]. Although TE and TM modes are treated separately, they may be coupled by the boundary conditions in an optical fibre to give the so called 'EH', 'HE' or 'LP' guided waves (Kapoor and Singh 2000). The second objective is to demonstrate that there are two interpretations of exactly the same electromagnetic fields. The fields are composed of elements on opposing branches of the relativistic light cones but the branches can be switched without altering the spatial boundary conditions and patterns of the modal fields.

In this work, the interactions between the field and dielectric are accounted for by a scalar frequency-dependent relative permittivity $\varepsilon_r = n^2$ (taking the relative permeability $\mu_r = 1$ for simplicity). The dielectric is assumed to have no optical loss or gain and the approximation is made that the energy is confined to a uniform refractive index $n$. A unit vector **u** defines the forward direction of propagation along Oz, the fibre's axis. The boundary conditions are not required in any detail so that Cartesian co-ordinates permit a straightforward matrix description of the local rotation of the 'X','Y' axes. The transverse vectors are now ordered as column matrices:

$$\mathbf{B_T} = \begin{bmatrix} B_x \\ B_y \end{bmatrix} \; ; \; \mathbf{E_T} = \begin{bmatrix} E_x \\ E_y \end{bmatrix} \; ; \; \nabla_\mathbf{T} = \begin{bmatrix} \partial_x \\ \partial_y \end{bmatrix} \qquad 2.1$$

$$(\mathbf{u} \times \mathbf{E_T}) = \begin{bmatrix} -E_y \\ E_x \end{bmatrix} = \sigma\mathbf{E_T} \quad ; \sigma = \begin{bmatrix} 0 & -1 \\ 1 & 0 \end{bmatrix} ; \qquad 2.2$$

$$(\mathbf{u} \times \nabla_\mathbf{T}) \rightarrow \begin{bmatrix} -\partial_y \\ \partial_x \end{bmatrix} = \sigma\nabla_\mathbf{T} \qquad 2.3$$

Note $\sigma^2 = -1$ but $\sigma^{tr}\sigma = 1$ where $^{tr}$ denotes row-column transposition. Rotations are kept as real matrix operations and $i$ is reserved solely for the phasor field variations $\exp i\,\zeta$ where $\zeta = (\omega t - kz + \phi)$. Hence $i\mathbf{E_T}$ has a $90^0$ phase shift while $\sigma\mathbf{E_T}$ has a $90^0$ rotation.

Vector and matrix equivalences can be established with $\nabla_\mathbf{T}.\mathbf{F_T} = \nabla_\mathbf{T}^{tr}\mathbf{F_T}$ for any transverse vector field $\mathbf{F_T}$. Given a matrix $\Theta = \exp(\sigma\theta) = (\cos\theta + \sigma\sin\theta)$, direct evaluation shows that $\Theta\,\mathbf{F_T}$ is rotated with respect to $\mathbf{F_T}$ by an angle $\theta$. Making use of $\sigma^{tr}\sigma = 1$, it is found that $\Theta^{tr}\Theta = 1$. In particular one may write for example:

$$(\Theta\sigma\nabla_\mathbf{T})^{tr}(\Theta\sigma\mathbf{F_T}) = \nabla_\mathbf{T}{}^{tr}\sigma^{tr}(\Theta^{tr}\Theta)\,\sigma\mathbf{F_T} = (\Theta\nabla_\mathbf{T})^{tr}(\Theta\mathbf{F_T}) = \nabla_\mathbf{T}{}^{tr}\,\mathbf{F_T} \qquad 2.4$$

This holds even if $\Theta$ varies with $(t,\ z)$ provided that $\Theta$ is independent of the transverse coordinates: $\nabla_\mathbf{T}(\Theta\,\mathbf{F_T}) = \Theta(\nabla_\mathbf{T}\,\mathbf{F_T})$. Appendix A gives Maxwellian TE fields ($E_z = 0$) from:

$$\nabla_\mathbf{T}{}^{tr}\, c\mathbf{B}_{\mathbf{T}\text{te}} + \partial_z(cB_z) = 0 \qquad 2.5$$

$$\nabla_\mathbf{T}{}^{tr}(\sigma\mathbf{E}_{\mathbf{T}\text{te}}) - (1/c)\,\partial_t(cB_z) = 0 \qquad 2.6$$

$$\partial_z(c\mathbf{B}_{\mathbf{T}\text{te}}) + (1/c)\,\partial_t[n^2(\sigma\,\mathbf{E}_{\mathbf{T}\text{te}})] - \nabla_\mathbf{T}\,(cB_z) = c\mu_0\mu_r\,\sigma\mathbf{J_T} \qquad 2.7$$

Appendix A gives the equivalent Maxwellian TM fields ($B_z = 0$) from:

$$(\sigma\nabla_\mathbf{T})^{tr}\ [n^2(\sigma\mathbf{E}_{\mathbf{T}\text{tm}})] + \partial_z(n^2E_z) = \mu_r\,\rho/\varepsilon_o \qquad 2.8$$

$$(\sigma\nabla_\mathbf{T})^{tr}(c\mathbf{B}_{\mathbf{T}\text{tm}}) - (1/c)\,\partial_t\,(n^2E_z) = -\,c\mu_0\mu_r J_z \qquad 2.9$$

$$\partial_z\ [(\sigma\mathbf{E}_{\mathbf{T}\text{tm}})] + (1/c)\,\partial_t(c\mathbf{B}_{\mathbf{T}\text{tm}}) - \sigma\nabla_\mathbf{T}\,(E_z) = 0 \qquad 2.10$$

Consider waves propagating as exp $i(\omega t - kz)$ where currents $\mathbf{J}$ and charge density $\rho$ are set to zero with $\mu_r$=1. The waves have a *transverse* propagation constant $\kappa$ where

$$\nabla_\mathbf{T}.\nabla_\mathbf{T}A_z = \nabla_\mathbf{T}{}^{tr}\nabla_\mathbf{T}A_z = (\partial_x{}^2 + \partial_y{}^2)\,A_z = -\kappa^2 A_z \quad \text{for } A_z = E_z \text{ or } B_z \qquad 2.11$$

The value of $\kappa$ will be the same for the TE and TM modes if these combine as a single EH or HE mode. From equs. 2.5-10, the TE and TM fields can then be determined from:

$$\sigma n\mathbf{E}_{\mathbf{T}\text{te}} = -i\,(n\omega/c)\,\nabla_\mathbf{T}(cB_z)/\kappa^2 \quad ; \quad c\mathbf{B}_{\mathbf{T}\text{te}} = -ik\,\nabla_\mathbf{T}(cB_z)/\kappa^2 \qquad 2.12$$

$$(\sigma n\mathbf{E}_{\mathbf{T}\text{tm}}) = -ik\,\sigma\nabla_\mathbf{T}(nE_z)/\kappa^2 \quad ; \quad c\mathbf{B}_{\mathbf{T}\text{tm}} = -\,i(n\,\omega/c)\ \sigma\nabla_\mathbf{T}(nE_z)/\kappa^2 \qquad 2.13$$

with

$$k^2 + \kappa^2 = n^2\omega^2/c^2 \qquad 2.14$$

Notice that $\sigma\mathbf{E}_{\mathbf{T}\text{te/tm}}$ and $c\mathbf{B}_{\mathbf{T}\text{te/tm}}$ are proportional to each other over the whole cross section.

The second part of this section considers the relativistic properties of these equations. For a Lorentz transformation along the Oz axis with a velocity $v$ where tanh $\alpha = v/c$, it is possible to divide every quantity into components (subscript +) that transform along the positive light cone and components (subscript $-$) that transform along the negative light cone. For example $2\,z = z_+ + z_-$ where $z_+ = (z - ct)$ and $z_- = (\,z + ct\,)$ where:

$$z_-' = z' + ct' = \exp(-\alpha)\,(\,z + ct) = \exp(-\alpha)\ z_- \ ; \qquad 2.15$$

$$z_+' = z' - ct' = \exp(\alpha)\,(\,z - ct) = \exp(+\alpha)\ z_+ \ ; \qquad 2.16$$

Define differential operators and transverse vectors on these different branches of the light cone from:

$$\partial_- = \partial_z - (1/c)\,\partial_t \ ; \ \partial_+ = \partial_z + (1/c)\,\partial_t\ ; \qquad 2.17$$

$$(c\mathbf{B}_{\mathbf{T}\text{te}} - \sigma\mathbf{E}_{\mathbf{T}\text{te}}) = \mathbf{F}_{\text{te}+} \ ; \ (c\mathbf{B}_{\mathbf{T}\text{te}} + \sigma\mathbf{E}_{\mathbf{T}\text{te}}) = \mathbf{F}_{\text{te}-}\ ; \qquad 2.18$$

$$(n^2\mathbf{E}_{\mathbf{T}\text{tm}} + \sigma c\mathbf{B}_{\mathbf{T}\text{tm}}) = \mathbf{F}_{\text{tm}+} \ ; \ (n^2\mathbf{E}_{\mathbf{T}\text{tm}} - \sigma c\mathbf{B}_{\mathbf{T}\text{tm}}) = \mathbf{F}_{\text{tm}-}\ ; \qquad 2.19$$

These definitions allow the TE equations to be written as

$$\nabla_T^{tr}\,\mathbf{F}_{te+} + \partial_+(cB_z) = 0\ ;\ \nabla_T^{tr}\,\mathbf{F}_{te-} + \partial_-(cB_z) = 0\ ; \qquad 2.20$$

$$\partial_+\{[1+\tfrac{1}{2}(n^2-1)]\mathbf{F}_{te-} - \tfrac{1}{2}(n^2-1)\,\mathbf{F}_{te+}\} + \partial_-\{[1+\tfrac{1}{2}(n^2-1)]\,\mathbf{F}_{te+} - \tfrac{1}{2}(n^2-1)\,\mathbf{F}_{te-}\} - 2\nabla_T(cB_z) = 0 \qquad 2.21$$

Similarly, after some algebraic manipulation, the TM equations can be written as:

$$\nabla_T^{tr}\,\mathbf{F}_{tm+} + \partial_+(n^2 E_z) = 0\ ;\ \nabla_T^{tr}\,\mathbf{F}_{tm-} + \partial_-(n^2 E_z) = 0\ ; \qquad 2.22$$

$$\partial_+\{[1+\tfrac{1}{2}(n^2-1)]\,\mathbf{F}_{tm-} - \tfrac{1}{2}(n^2-1)\mathbf{F}_{tm+}\} + \partial_-\{[1+\tfrac{1}{2}(n^2-1)]\mathbf{F}_{tm+} - \tfrac{1}{2}(n^2-1)\,\mathbf{F}_{tm-}\} - 2\nabla_T(n^2 E_z) = 0 \qquad 2.23$$

The similarity of the equs. 2.20-21 and 2.22-23 demonstrates the similarity of the TE and TM fields though with the former 'driven' by $cB_z$ and latter 'driven' by $n^2 E_z$.

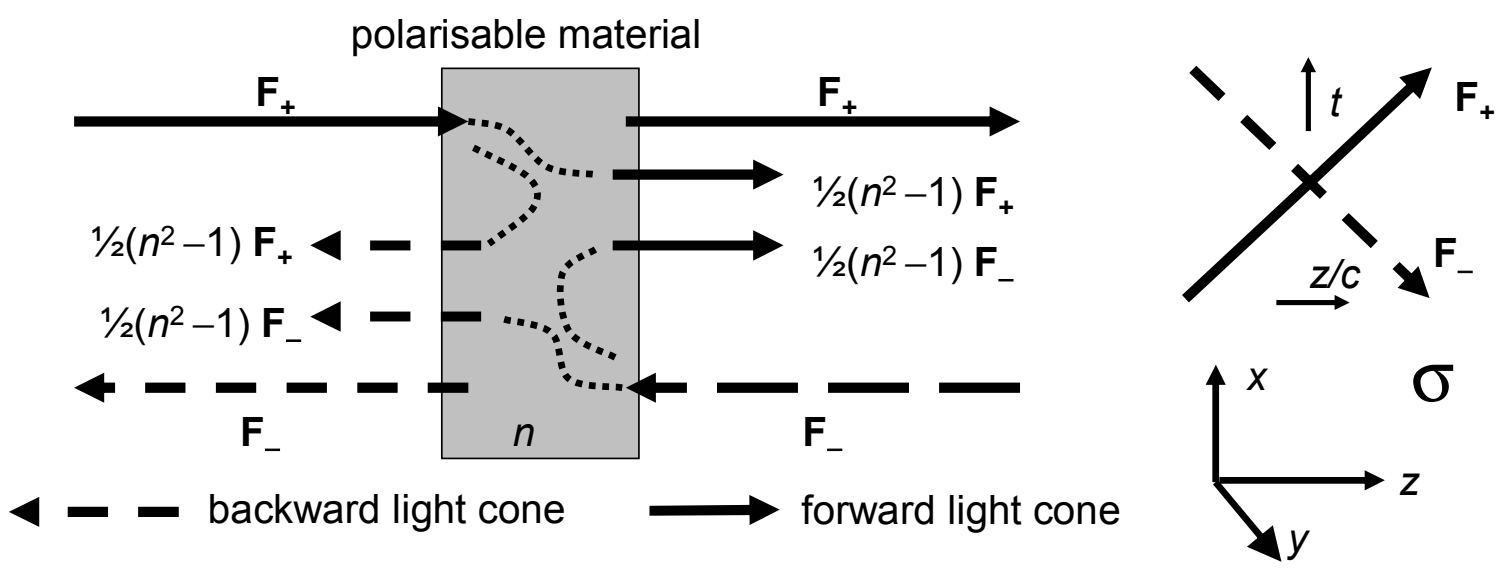

**Figure 1:** *Scattering of electromagnetic fields on different branches of the light cone*

Figure 1 sketches how fields $\mathbf{F}_+$ on the forward light cone are scattered by the polarisable material into $\tfrac{1}{2}(n^2-1)\,\mathbf{F}_+$ still on the forward light cone but $\tfrac{1}{2}(n^2-1)\,\mathbf{F}_+$ is scattered onto the backward light cone. Similarly the fields $\mathbf{F}_-$ on the backward light cone are scattered by the polarisable material into $\tfrac{1}{2}(n^2-1)\,\mathbf{F}_-$ still on the backward light cone but $\tfrac{1}{2}(n^2-1)\,\mathbf{F}_-$ is scattered onto the forward light cone. Hence the polarisable material couples fields on the forward and backward branches of the light cone.

On changing the sign of $\sigma$ in the equs. 2.18 and 2.19 one notes that the fields on the two branches of the light cone are interchanged: fields designated as $\mathbf{F}_+$ and $\mathbf{F}_-$ change into $\mathbf{F}_-$ and $\mathbf{F}_+$ respectively for both TE and TM modes . Changing the sign of $\sigma$ essentially interchanges the forward and backward light cones (Figure 2). Now equs. 2.20-23 show that, provided one simultaneously interchanges $\partial_+$ and $\partial_-$ into $\partial_-$ and $\partial_+$ respectively so that the sign of $t$ changes , then Maxwell's equations remain unaltered.

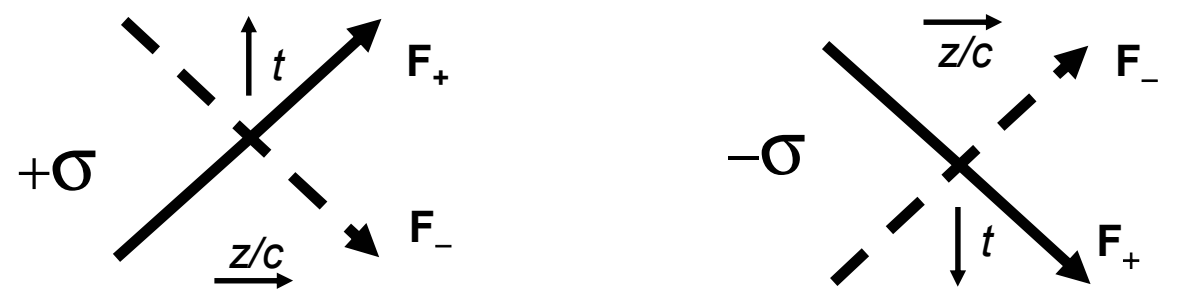

**Figure 2:** *Schematic: reversal of σ interchanges forward and backward light cones*

Consequently, all solutions of Maxwell's equations have two relativistic interpretations. In the first, conventional, interpretation, the fields have time travelling forward with relevant components on the two branches of the light cone. The second, unconventional interpretation is that the fields have time travelling backwards with the fields on the two light cone branches interchanged and the sign of $\sigma$ is changed. Changing the sign of $\sigma$ changes the chirality i.e. Ox,Oy,Oz becomes left handed instead of right handed. Normally this left-handed solution is dismissed as unphysical but this is reconsidered in section 5.

## 3. A new family of Maxwellian solutions with local helical rotation of modal fields.

Suppose that, relative to a fixed frame of reference *OxOyOz*, there are TE fields $\mathbf{E}_{\mathrm{Tte}}$ and $c\mathbf{B}_{\mathrm{Tte}}$ in the *OxOy* plane varying as exp $i(\omega t - kz)$. It is not obvious that these transverse fields can have an additional *local helical* rotation (Figure 3) but still retain the same modal solutions with frequency $\omega$ axial propagation constant $k$ and transverse propagation constant $\kappa$.

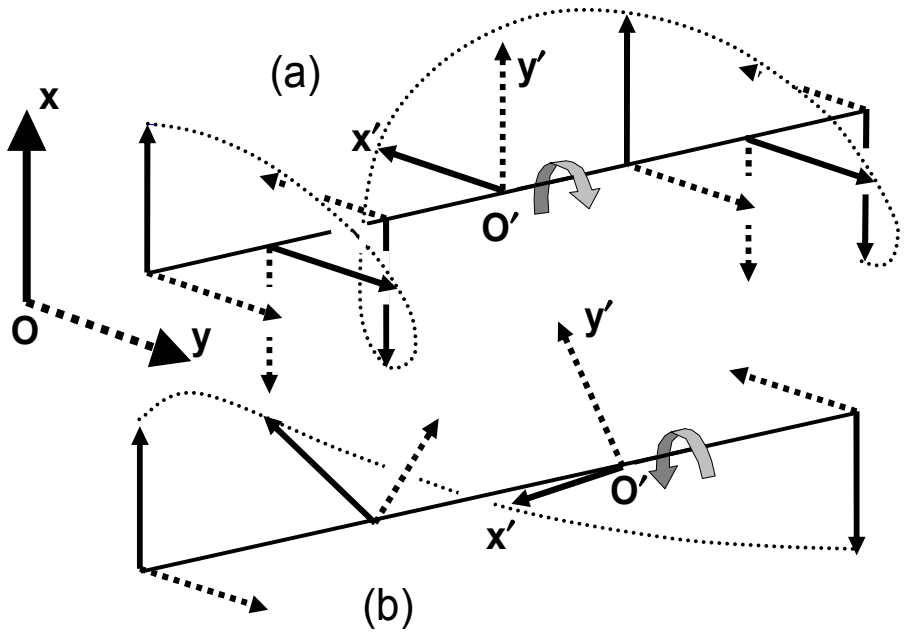

**Figure 3.**
*Schematic helical rotation of reference axes Ox Oy about Oz into transverse axes Ox' Oy'.*
Angle of rotation depends on position $z$.
(a) Positive rotation: (b) Negative rotation.

However consider a helical rotation operator $\Theta$ where

$$\Theta = \exp \sigma\, \Omega\, (t - z/v_{\mathrm{h}}) \qquad 3.1$$

The helical angular frequency $\Omega$ may at present be chosen arbitrarily but the velocity $v_{\mathrm{h}}$ will be specified shortly. Retaining the same values for $z$ and $t$, the unit transverse directions $\underline{\mathbf{X}}$, $\underline{\mathbf{Y}}$ are helically rotated at each point $(x,y)$ into $\underline{\mathbf{X}}'$, $\underline{\mathbf{Y}}'$. So that any transverse vector $\mathbf{F}_{\mathbf{T}}$ is then helically rotated into $\mathbf{F}_{\mathbf{T}}' = \Theta\, \mathbf{F}_{\mathbf{T}}$ or $\mathbf{F}_{\mathbf{T}} = \Theta^{-1}\, \mathbf{F}_{\mathbf{T}}'$. We shall consider boundary conditions later. With primes placed on the twisted fields one then seeks solutions to Maxwell's TE equations where , using equ. 3.1:

$$\mathbf{E}_{\mathrm{Tte}} = (\Theta^{-1}\mathbf{E}_{\mathrm{Tte}}')\,;\; c\mathbf{B}_{\mathrm{Tte}} = (\Theta^{-1}\, c\mathbf{B}_{\mathrm{Tte}}')\,;\; \nabla_{\mathbf{T}} = \Theta^{-1}\, \nabla_{\mathbf{T}}' \qquad 3.2$$

It is emphasised that this is a local rotation of the transverse axes and not the total rotation considered by Modesitt (1970). It could be argued that the $\Theta$ operation creates a spin rotation rather than the orbital rotation occurring in the Laguerre-Gaussian modes . The localisation of the rotation of the transverse axes at each transverse point means that the vector field $B_z'$ is the same as $B_z$ . The coordinates $t$ and $z$, along with their origins, do not change. Now re-write TE equs. 2.5 -2.7 in terms of $\mathbf{E}_{\mathrm{Tte}}'$, $c\mathbf{B}_{\mathrm{Tte}}'$ and $\nabla_{\mathbf{T}}'$ using equ. 3.2:

$$(\Theta^{-1}\, \nabla_{\mathbf{T}}')^{\mathrm{tr}}.(\Theta^{-1}c\mathbf{B}_{\mathrm{Tte}}') + \partial_z (cB_z') = 0\,; \qquad 3.3$$

$$(\Theta^{-1}\nabla_{\mathbf{T}}')^{\mathrm{tr}}.(\Theta^{-1}\, \sigma\mathbf{E}_{\mathrm{Tte}}') - (1/c)\, \partial_t (cB_z') = 0\,; \qquad 3.4$$

$$\partial_z (\Theta^{-1}c\mathbf{B}_{\mathrm{Tte}}') + (n^2/c)\, \partial_t (\sigma\, \Theta^{-1}\mathbf{E}_{\mathrm{Tte}}') - \Theta^{-1}\, \nabla_{\mathbf{T}}' (cB_z') = 0\,; \qquad 3.5$$

Now, using equ. 2.4, $\Theta$ can be eliminated from equs. 3.3- 4 leaving equs. 2.5-6 but in terms of the twisted primed fields. The operator $\Theta^{-1}$ can only be cancelled from equ. 3.5 provided that one satisfies:

$$(\partial_z \Theta^{-1})\, c\mathbf{B}_{\mathrm{Tte}}' + (n^2/c)\, (\partial_t \Theta^{-1})\, (\sigma\mathbf{E}_{\mathrm{Tte}}') = 0 \quad ; \qquad 3.6$$

This leaves equ. 2.7 but with primed fields:

$$\partial_z(c\mathbf{B}_{\mathbf{T}te}') + (n^2/c)\,\partial_t(\sigma\,\mathbf{E}_{\mathbf{T}te}') - \nabla_{\mathbf{T}}\,(cB_z') = 0\,. \qquad 3.7$$

Now given equ. 3.1; $(1/c)\partial_t\Theta^{-1} : \partial_z\Theta^{-1} = \sigma(\Omega/c) : -\sigma(\Omega/v_h)$ then equ. 2.12 combined with equ. 3.6 shows that the helical rotation frequency $\Omega$ takes arbitrary values provided that the helical velocity satisfies:

$$v_h = c^2k/(n^2\omega) = d\omega/dk = v_g\,. \qquad 3.8$$

where $v_g$ is the group velocity $d\omega/dk$ calculated from equ. 2.14 with a constant $\kappa$. With a permittivity that is weakly dependent on frequency, methods used by Haus (1995) and Bers(2000) can confirm that $v_h = v_g$ is still required.

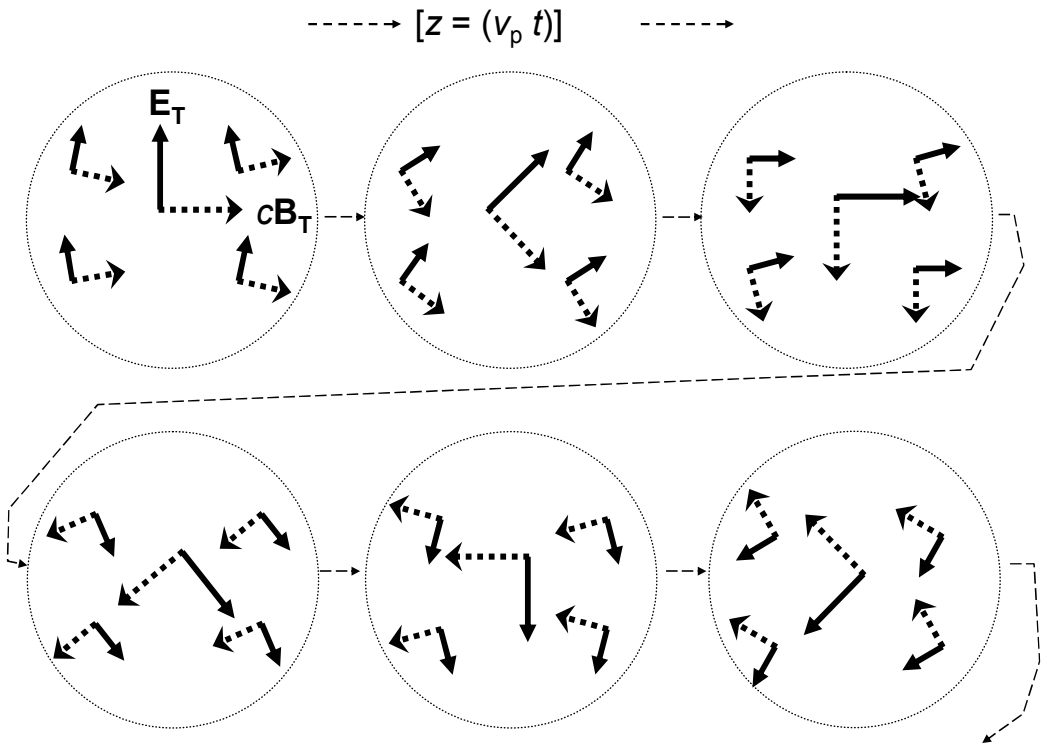

**Figure 4.** *Schematic local helical rotation of transverse fields.*

The diagrams show a linearly polarised $\mathbf{E_T}$ (solid line) and $c\mathbf{B_T}$ (dotted line) as the observer *travels with the phase velocity*. The helical rotation (*travelling at the group velocity*) of all transverse vectors makes the fields appear to rotate locally about their local $(x,y)$ positions.

The similar formalisms of both TM and TE equations can show that similar results are found for TM modes. As a consequence, for any classical mode with frequency $\omega$ and propagation constant $k$ with transverse fields $\mathbf{F_T}$ there is a family of modes with a local helical rotation of all transverse vectors: $\mathbf{F_T}' = \Theta\,\mathbf{F_T}$ where $\Theta = \exp[\sigma\Omega(t - z/v_g)]$. This local helical rotation can have an arbitrary helical rotational frequency $\Omega$ but must have a helical velocity equal to the modal group velocity in order for the primed fields to satisfy the same Maxwellian equation with the same $\omega$ and $k$. The boundary conditions will be discussed later in sec.4. Note again that, this local helical rotation with $\Omega$ different from $\omega$ is not the same as orbital rotation found in the single frequency Laguerre-Gaussian modes.

Because $\Theta^{tr}\Theta = 1$, this local helical rotation $\Theta$ of the fields has no effect on the *classical* calculation of energy averaged over the cross-section of the guide (denoted by $\langle\ \rangle$):

$$\tfrac{1}{2}\langle\, (\Theta\mathbf{E_T})^{tr}(\Theta\mathbf{D_T}) + (\Theta\mathbf{H_T})^{tr}(\Theta\mathbf{B_T}) + E_z D_z + H_z B_z \,\rangle$$
$$= \tfrac{1}{2}\langle\, \mathbf{E_T}^{\mathbf{tr}}\mathbf{D_T} + \mathbf{H_T}^{\mathbf{tr}}\mathbf{B_T} + E_z D_z + H_z B_z \,\rangle \qquad 3.9$$

**4. Counter helical modulation forming a wave-packet of a classical mode**

Now notice from equs. 2.12-13 that all the transverse fields are proportional to $\nabla\varphi$ where $\varphi$ is proportional to $E_z$ and/or $B_z$ in appropriate combinations. The transverse fields of any classical Maxwellian mode can be said to driven by some $\nabla\varphi$. Now the transverse fields, $\sigma\mathrm{E_T}$ or $\mathbf{B_T}$, can be generically described by a vector $\mathbf{F_T}$. With a helical modulation as discussed in section 3, the rotated fields become $\mathbf{F_{TR}} = \exp(\sigma\,\Omega\,\tau)\,\nabla\varphi$. Then consider an associated field $\mathbf{F_{TA}} = \exp(\sigma\ \phi)\exp(-\sigma\,\Omega\,\tau)\,\nabla\varphi$ oriented at an angle $\phi$ with respect to $\mathbf{F_{TR}}$ and helically modulated with an opposing helicity. These fields are assumed to combine over a region $-\tau_1 \le \tau \le \tau_2$ where $\tau = (t - z/v_h)$ so as to give net fields $\mathbf{\Phi_\Omega}$ :

$$\mathbf{\Phi_\Omega} = [\exp(\sigma\,\Omega\,\tau)\,\mathbf{F_{TR}} + \exp(-\sigma\,\Omega\,\tau)\,\mathbf{F_{TA}}]_{(-\tau_1 \le \tau \le \tau_2)} \qquad 4.1$$

$\mathbf{\Phi_\Omega} = \exp(\sigma \tfrac{1}{2}\phi)\ [\cos(\Omega\tau - \tfrac{1}{2}\phi)]\ \nabla\varphi_{(-\tau_1 \le \tau \le \tau_2)}$ 4.2

With the combined counter-helical rotation the resulting fields are simple linear modulations of the underlying mode that matches all the boundary conditions of the waveguide so that the boundary conditions are still met. The polarisation is determined by the temporal phase differences between $\sigma\mathbf{E_T}$ and $c\mathbf{B_T}$ and these relative phases are not altered by the rotating helical modulation so that the optical polarisation [Hecht 1974] is unchanged. For example $\nabla\varphi$ can be a combination of fields with arbitrary elliptical polarisation and the combined fields $\mathbf{\Phi_\Omega}$ retain this polarisation of the underlying mode.

Now travelling at the group velocity ($v_g = v_h$) with $\tau = (t - z/v_h)$, suppose that these combined fields $\mathbf{\Phi_\Omega}$ vanish at 'boundaries' $\tau = -\tau_1$ and $\tau = \tau_2$ so that at these values $\cos(\Omega\tau - \tfrac{1}{2}\phi) = 0$ yielding arbitrary integers M and N:

$(-\Omega\tau_1 - \tfrac{1}{2}\phi) = (-M - \tfrac{1}{2})\pi\ ;\ (\Omega\tau_2 - \tfrac{1}{2}\phi) = (N - \tfrac{1}{2})\pi$ 4.2

$\Omega(\tau_1 + \tau_2) = (M + N)\pi\ ;$ 4.3

$\Omega(\tau_2 - \tau_1) + \phi = (N - M - 1)\pi;$ 4.4

This leads to two distinct families of solutions: one family with N = M+1 and the other with M = N. The packet duration is measured by the 'time of flight' $\tau_0 = (\tau_1 + \tau_2)$.

The family where N = M yields $\Omega\tau_0 = 2M\pi$ so that for this family the shortest packet length is zero: $\tau_0 = 0$ when M = 0. Because the minimum packet is zero (i.e. no packet at all), this family is discarded from further consideration.

The family where N = M+1 has a shortest value of $\tau_o$ when M = 0 and N = 1 giving $\Omega\tau_o = \pi$. Now postulate that packet $\mathbf{\Phi_\Omega}$ should always trap one whole (†) temporal period of the underlying mode that varies as $\cos(\omega t - kz + \xi)$ with some arbitrary phase $\xi$. This requires that $\omega\tau_o = 2\pi$ so that this family of solutions where N = M+1 gives helical modulation frequencies identical to the Schrödinger frequencies of the quantum oscillator:

$\Omega = (2M+1)\omega/2$ 4.5

[† If one chose to trap a half period then, when $\Omega = \omega$, there is an inconsistency by forming a zero frequency field.]

Figure 5 sketches how three lowest order wave-packets of this family might appear with a linearly polarised modal field determined from $\nabla\varphi$ travelling as $\cos[\omega(t - z/v_p)]$ but modulated as in equ.4.1 with $\Omega = 1/2\omega$ [ Fig.4 (a)]; $\Omega = 3/2\omega$ ]Fig.4(b)]; $= 5/2\omega$ [Fig.4(c)]; and with all three packets superimposed in Fig.4 (d). The phase factor $\phi$ (equ. 4.4) alters the position of the centre of the wave-packet but the general features sketched in Figure 5 do not depend on $\phi$. The group velocity $v_g$ is the velocity of the envelope as determined by the counter helical rotations travelling with the helical velocity $v_h = v_g$. The phase velocity is determined by the underlying mode varying as $\exp i[\omega(t - z/v_p) + \xi]$ ; ($v_p = \omega/k$) .

The *classical* field energy can be found through space-time averages of the square of the fields in the usual classical manner for evaluating field energies (see equ. 3.9) :

$\langle\langle \mathbf{\Phi_\Omega}^{tr*}\mathbf{\Phi_\Omega} \rangle\rangle = \langle\langle \mathbf{F_R}^* \mathbf{F_R} \rangle\rangle + \langle\langle \mathbf{F_A}^* \mathbf{F_A} \rangle\rangle$ 4.6

This energy is the same as if the two fields had their two separate classical energies added together. The wave-packet created by the two *counter helically rotating* waves then does not change either the type of polarisation or the net classical energy.

One of the proposals in this work is that wave-packets, $\mathbf{\Phi}_{\Omega}$, create the true observable quantities and form the proposed classical model for a photon. While the group velocity changes with the modal frequency ω, the helical velocity and hence the packet velocity do not change with changes of the helical frequency Ω for a fixed ω. However deviations between $v_h$ and $v_g$ would lead to the break-up of such a wave-packet unless some new effect can be found to preserve the integrity. This new effect is addressed in the next section.

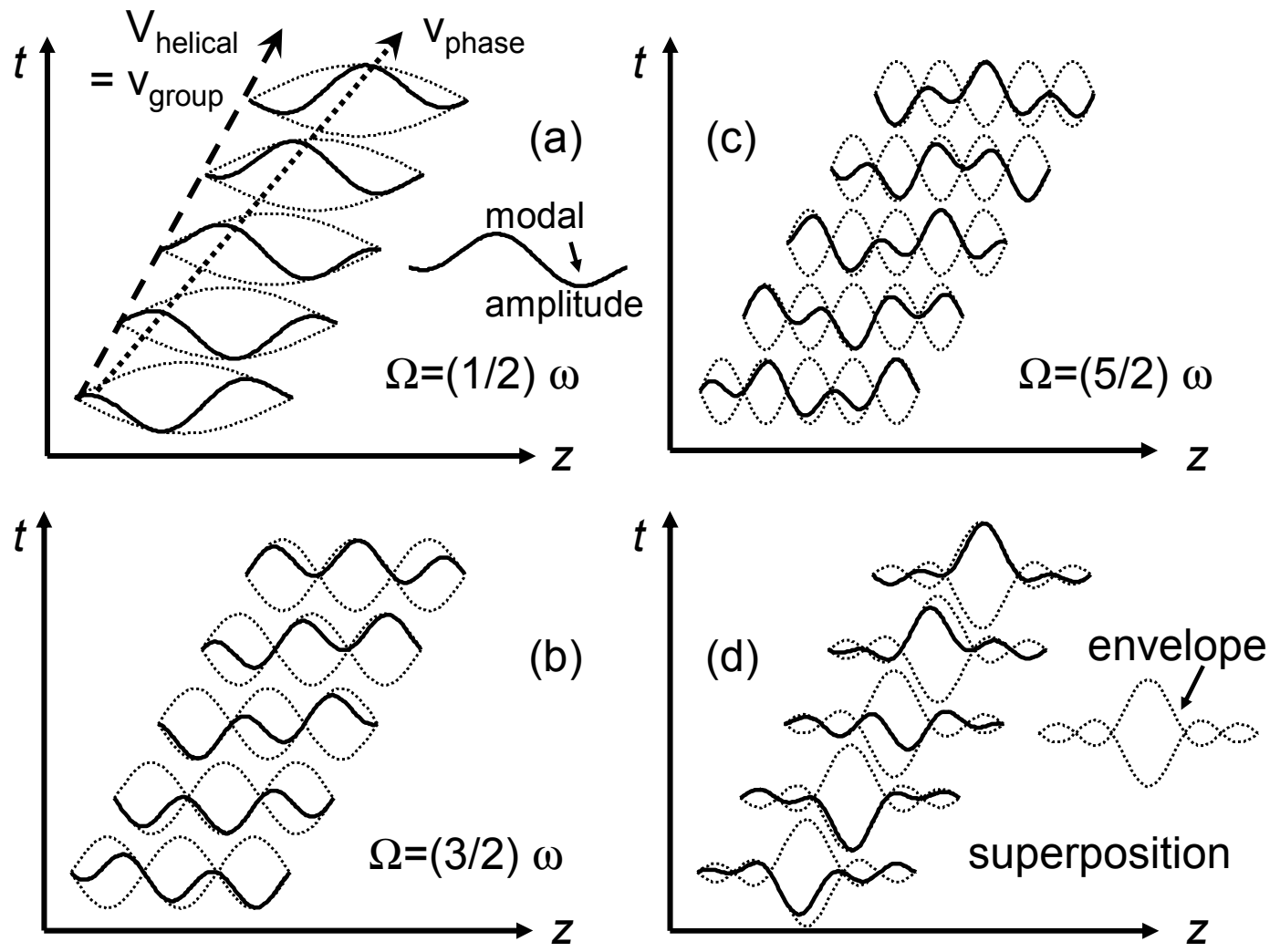

**Figure 5** *Schematic for counter rotating helically modulated modal wave-packets* with a linearly polarised underlying mode. Dashed lines show the envelope of the modulation; solid lines show the resulting wave amplitudes of the modulated mode. The envelope travels with the group velocity while the peak of the underlying modal amplitude travels with the phase velocity inside the envelope.

## 5. Advanced and retarded waves: PRAHM modes

When Jackson (1962) or Wheeler and Feynman (1945) discuss *retarded* waves they consider waves varying as $\exp[i\omega(t - r/c)]$ radiating energy outwards away from the source at $r = 0$. Their *advanced* waves vary as $\exp[i\omega(t + r/c)]$ and bring energy inwards towards the source at $r = 0$. Jackson discards advanced solutions as unphysical while Wheeler and Feynman considered the effects of advanced waves on radiation reaction. A common view of advanced waves is that these must be dismissed because, at least by themselves, they appear to violate causality by bringing energy from the future into the past or present. However this view is modified in this paper.

In this work, the conventional retarded wave varies as $\exp i(\omega t - kz)$ along the guide and, as usual, carries power generated in the past or present into the future. The retarded fields have components $\mathbf{F}_{+}$ and $\mathbf{F}_{-}$ (see equs. 2.18, 2.19) on the forward and backward branches of the light cone. The unconventional advanced wave is now defined by interchanging these two branches as described in section 2. Both $t$ and ω change sign so that advanced waves remain in synchronism with retarded waves varying as $\exp i(\omega t - kz)$. However because the advanced wave has component fields lying on opposite light cones to that of the retarded wave, the energy is now propagating in the backward direction. Advanced waves by definition are excited in the future and consequently advanced waves cannot be modulated or altered in the present without violating causality. Consequently their existence has to rely on random disturbances or 'noise' without cause in the present.

This difference of the direction of power flow between the advanced and retarded wave has a dramatic effect on the physics of the wave-packet discussed in section 4, as sketched in Figure 6. Taking equ. 4.1 trapping one period of a classical mode of frequency $\omega$, then we should now write with subscript R/A indicating retarded/advanced respectively:

$$\mathbf{\Phi}_{\mathbf{M}} = [\exp(\sigma\,\Omega\,\tau)\,\nabla\varphi_R + \exp(-\sigma\,\Omega\,\tau)\exp(\sigma\ \phi)\ \nabla\varphi_A\,]_{(-\tau_1 \le \tau \le \tau_2)} \qquad 5.1$$

$$\Omega = (M+½)\omega$$

The gradient functions generating the advanced and retarded fields have equal amplitudes: $\nabla\varphi_R = \nabla\varphi_A$ and both vary with identical phase velocities as cos $[\omega t - kz + \phi]$. However, while the retarded wave drives power towards the wave front of the packet at $\tau = (t - z/v_g) = \tau_2$ , the advanced wave provides feedback of the power back to the wave-rear at $\tau = -\tau_1$. It is proposed that energy is carried by the waves only in the region of overlap (Figs. 5 and 6) $(-\tau_1 \le \tau \le \tau_2)$. This wave-packet trapping one temporal period of the underlying classical mode has a resonant modulation frequency $\Omega = (M + ½)\omega$ for integer M: the Schrödinger frequencies of standard quantum theory. This new type of resonant wave-packet is referred to here as a PRAHM mode (**P**acket of **R**etarded and **A**dvanced **H**elically **M**odulated mode). 'Ein Prahm' is a term for a flat bottomed ferry or barge in German so that a PRAHM mode is seen as ferrying packets of electromagnetic energy.

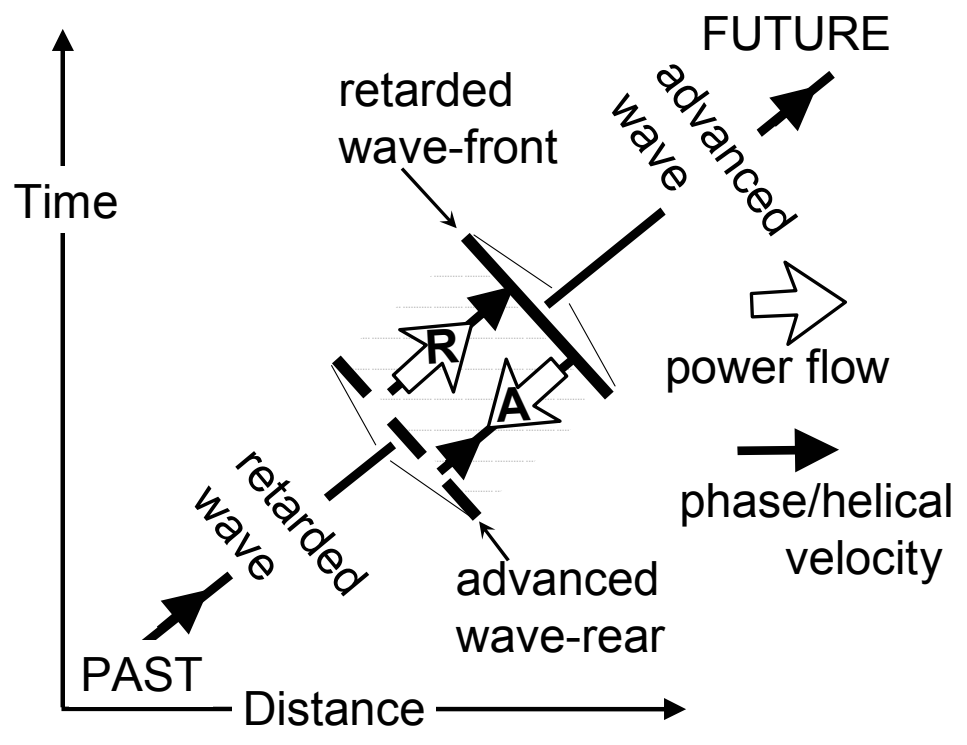

**Figure 6**
*Schematic formation of PRAHM mode:* Packets of Retarded and Advanced Helically Modulated modes creating a resonance. The advanced wave feeds back power to maintain resonant state with retarded wave carrying power forward. The packet travels at the group velocity.

An alternative argument that reaches the same conclusions is to reverse $\sigma$ and reverse **E**. The Maxwell equs. of 2.20-2.23 are found to be unaltered but the reversal of **E** must reverse the direction of the Poynting vector $\mathbf{E} \times \mathbf{H}$ into $-(\mathbf{E} \times \mathbf{H})$. The right handed system of fields and propagation vector $\{\, \Theta\, \mathbf{E}_{\mathbf{TR}}\, ,\ \Theta\, \mathbf{B}_{\mathbf{TR}},\ \mathbf{k} \,\}$ of the retarded waves is changed into $\{\, -\Theta^{-1}\, \mathbf{E}_{\mathbf{TA}}\, ,\ \Theta^{-1}\, \mathbf{B}_{\mathbf{TA}},\ \mathbf{k} \,\}$ being a right handed system for the advanced waves or equally $\{\, \Theta^{-1}\, \mathbf{E}_{\mathbf{TA}}\, ,\ \Theta^{-1}\, \mathbf{B}_{\mathbf{TA}},\ \mathbf{k} \,\}$ is now a left-handed system. Left-handed waves were considered by Veselago (1968) and have created considerable interest. A key difference is that Veselago's left-handed waves are confined to special (left-handed) material where a change in the sign for $\varepsilon_r$ and $\mu_r$ has been engineered. Within limited volumes of such left-handed material, analysis shows that causality is not violated. In this work, it is proposed that energy is contained only in the overlap region of advanced and retarded waves. Hence causality is never violated by advanced waves acting alone.

## 6. Interaction energy in a PRAHM mode: a classical 'Planck' theorem

The concept that the PRAHM mode is resonant is a key feature of the theory. It is known that for a given source, only a certain amount of energy can be transferred into a single lossless resonance (Carroll 2000). A heuristic version of this reference can give a classical explanation of both Planck's law and constant. Consider a resonant wavelength of perfectly lossless transmission line with a characteristic impedance $Z_o$ that is driven at its input by a current source of $I \cos \omega t$. The line is shorted at its far end. With no initial fields, the current drives an average power $\frac{1}{2}I^2Z_o$ into the system (Fig.7.a). The short circuit reflects power after a time $\tau_0 = 2\pi/\omega$ (Fig.7.b,c). At time $2\tau_o$ the reflected fields return to the source and the impedance, seen by the source, changes from $Z_o$ into the reflected short circuit (Fig.7.d). All further power transfer ceases trapping within the lossless circuit a net energy $U_o = \frac{1}{2}I^2Z_o4\pi/\omega$. Now link the current source $I \cos \omega t$ to an oscillating electronic charge $e$ : $I = \zeta\, \omega\, e$. Taking, for example, $Z_o = 377\Omega$, the characteristic impedance of free space, then the energy transfer from the source to the lossless resonance is given by:

$$U_o = \xi\, hf \quad ; \qquad \xi \sim 0.6\, \zeta^2 \qquad 6.1$$

where h is Planck's constant. Of course the 'adjustment' factor $\xi$ cannot be rigorously evaluated by such an analysis but it shows why resonance is considered significant.

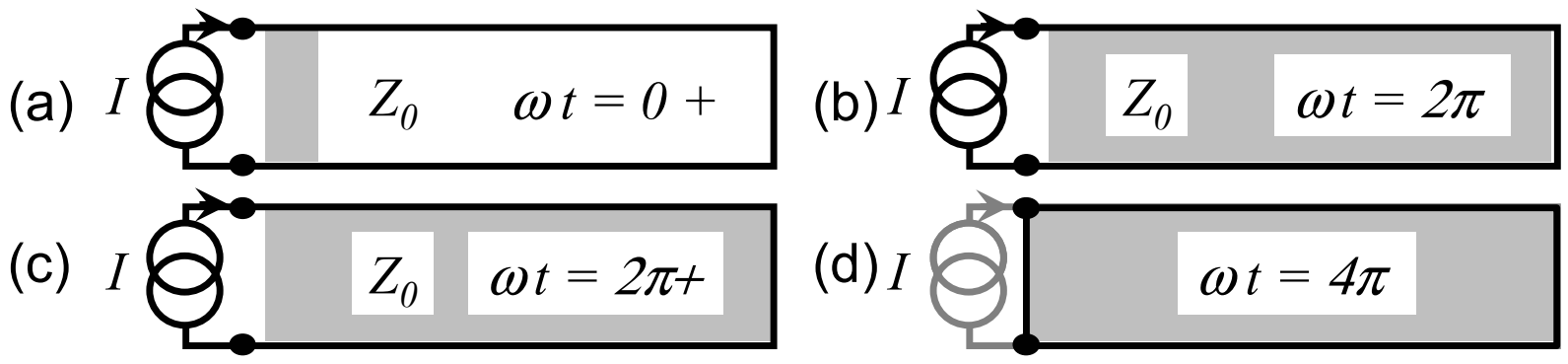

**Figure 7** *Energy transfer from current source into a resonant length of transmission line*

The idea that the energy is stored *within* the resonant system can also be used to support a further Planck-like theory. In this section it will be demonstrated how a helical energy that is proportional to the helical modulation frequency $\Omega = (N+\frac{1}{2})\omega$ can be attached to each PRAHM mode. However, it is helpful first to rehearse the arguments leading to the complex Poynting power flow theorem. Consider a current density $\mathbf{J^*}$ where * is the usual complex conjugation, reversing the sign of $i$ in the phasor temporal variations $\exp(i\omega t)$. Maxwell's equations give:

$$\mathbf{J} = -\partial_t\, \varepsilon_0\varepsilon_r\mathbf{E} + \mathrm{curl}\ \mathbf{H} \qquad 6.2$$

The energy per unit volume created the current $\mathbf{J^*}$working on the electric field $\mathbf{E}$ is then

$$\mathbf{E.J^*} = \mathbf{E.}\partial_t\, \varepsilon_0\varepsilon_r\mathbf{E^*} + \mathbf{E.}\mathrm{curl}\ \mathbf{H^*} \qquad 6.3$$

By using the symmetry of Maxwell's equations to evaluate $\mathbf{E.}\mathrm{curl}\ \mathbf{H^*}$ it is possible to demonstrate the complex Poynting power flow theorem [for example Sander and Reed 1986 ] which reduces to [ $\langle\ \rangle$ implying integration over cross section]:

$$\langle\, \partial_z(\mathbf{E} \times \mathbf{H^*}) + i\,\omega\, [\varepsilon_0\varepsilon_r\mathbf{E^*.E} - \mu_0\mu_r\mathbf{H^*.H}]\, \rangle = \langle\, \mathbf{E}\,.\,\mathbf{J^*}\, \rangle \qquad 6.4$$

In regions of zero current density, the divergence of the Poynting vector is zero because in any single mode there is an equal amount of stored electric and magnetic energy.

Now one may re-structure this theory to find an energy of interaction between the advanced and retarded fields of the PRAHM modes that were outlined in section 5. Begin by interpreting equation 6.1 for the advanced fields as providing an effective advanced current:

$$\mathbf{J_{Aeff}} = -\partial_t\, \varepsilon_0\varepsilon_r\mathbf{E_A} + \mathrm{curl}\ \mathbf{H_A} \qquad 6.5$$

The helical rotation requires that, just as the advanced fields are helically rotating, the effective advanced current density is also helically rotating : exp $[-\sigma(\Omega\tau)]\ \mathbf{J}_{\mathbf{effA}}$ where, as usual, $\tau = (t - z/v_g)$. In order to establish the resonant wave-packet, the retarded electric field, exp $[\sigma(\Omega\tau)]\ \mathbf{E_R}$ carrying energy forward, has to work against this effective advanced current carrying energy backwards. Now evaluate the energy exchange between the helical modulation fields over an increment $\Delta\tau$ in the retarded time of flight giving:

$$\Delta\{\exp[\sigma(\Omega\tau)]\ \mathbf{E_R}\} = \{\exp[\sigma(\Omega\Delta\tau)]\ \mathbf{E_R}\} \qquad 6.6$$

The advanced waves carry power in the opposite direction to the retarded fields and consequently a positive increment in the *retarded* time of flight must be turned into a negative increment in the *advanced* time of flight when evaluating the work done by $\mathbf{E_R}$ against $\mathbf{J_A}$. The power flow analysis then evaluates an average over the cross section:

$$\text{Power flow per unit length} = \langle\ \{\exp[(-\sigma)\Omega(-\Delta\tau)]\ \mathbf{J}_{\mathbf{effA}}\}^{tr}\{\exp[\sigma(\Omega\Delta\tau)]\ \mathbf{E_R}\}\ \rangle \qquad 6.7$$

From this concept, appendix B demonstrates a novel interaction energy theorem:

$$\langle\langle -\partial_z[\ \mathbf{E_{TR}}^{tr}\sigma\ \ \mathbf{H_{TA}}^*] + \Omega\,[\mathbf{D_{TR}}^{tr}\sigma\mathbf{E_{TA}}^* - \mathbf{H_{TR}}^{tr}\sigma\mathbf{B_{TA}}^*]\rangle\rangle$$
$$= \langle\langle\ \mathbf{E_{TR}}^{tr}\mathbf{J_{TA}}^* + (E_{zR})\,J_{zA}^*\rangle\rangle \qquad 6.8$$

where $\langle\langle\ \rangle\rangle$ denotes averaging over the cross section of the wave-guide and over one temporal period. Here * denotes that increments of $-\Delta\tau$ for the advanced wave are matched to increments of $+\Delta\tau$ for the retarded waves. At the end of an interaction time $\Delta\tau$ that establishes the PRAHM mode, all effective current densities fall to zero so that, similar to the complex power flow theorem one now has a helical power flow theorem:

$$\langle\langle\ \Delta[\ \mathbf{E_{TR}}^{tr}\sigma\ \ \mathbf{H_{TA}}^*]\ \rangle\rangle = \langle\langle\Omega\ [\mathbf{D_{TR}}^{tr}\sigma\mathbf{E_{TA}}^* -\ \mathbf{H_{TR}}^{tr}\sigma\mathbf{B_{TA}}^*]\ \mathrm{v_g}\ \Delta\tau\rangle\rangle \qquad 6.9$$

At the end of section 5 it was pointed out that the advanced fields associated with the retarded fields $\{\mathbf{E_{TR}}, c\mathbf{B_{TR}}\}$ can be linked by changing the sign of $\mathbf{E}$:

$$\{\mathbf{E_{TA}}^*, c\mathbf{B_{TA}}^*\} \rightarrow \{-\mathbf{E_{TR}}, c\mathbf{B_{TR}}\} \qquad 6.10$$

However, substitution of 6.10 into 6.9 gives zero energy because $\mathbf{F_{TR}}^{tr}\,\boldsymbol{\sigma}\,\mathbf{F_{TR}} = 0$ for all transverse fields. On the other hand if the advanced fields are rotated with respect to the retarded fields by a fixed angle $\phi = \pi/2$ as in equation 5.1 then such a simple rotation cannot alter their advanced nature of the fields but does change the values to give:

$$\{\mathbf{E_{TA}}^*, c\mathbf{B_{TA}}^*\} \rightarrow \{-\sigma\mathbf{E_{TR}}, c\sigma\mathbf{B_{TR}}\} \qquad 6.11$$

On substitution of equ. 6.11 into 6.9 and considering a mode where $\Omega = (M+½)\omega$

$$\langle\langle\ \Delta[\ \mathbf{E_{TR}}^{tr}\sigma\ \ \mathbf{H_{TA}}^*)]\ \rangle\rangle = \langle\langle(M+½)\ [\mathbf{D_{TR}}^{tr}\mathbf{E_{TR}} +\ (\mathbf{H_{TR}}^{tr}\mathbf{B_{TR}})]\ v_g\ 2\pi\rangle\rangle \qquad 6.12$$

This is then interpreted as:

Interaction energy required to establish the PRAHM mode $\boldsymbol{\Phi}_\mathbf{M}$
$\propto$ (M+½) × (Classical Energy in retarded fields within packet) 6.13

Energy is then changed in units of the 'ground state' classical energy within the PRAHM mode as M changes its integer value. This result combined with equ.6.1 indicates a classical version of Planck's law but further work is required to understand how to transfer energy from an oscillating electron to the helical modulation in order to evaluate Planck's constant.

## 7. Promotion and demotion

As a result of the work in Section 6, promotion / demotion of energy in the PRAHM mode has a clear physical meaning of increasing / decreasing the helical modulation frequency by $\omega$. The task in this section is to complete the mathematical connection with the promotion / demotion operators of standard quantum theory.

The model here assumes that interaction with the electron helically modulates the *retarded* electromagnetic fields. The advanced fields cannot be modulated in the present because, by definition, advanced fields are generated in the future. Appropriate advanced fields with the correct frequency, polarisation and phase are postulated to be picked up by the retarded fields from a random background sea of electromagnetic advanced fields (forming noise) when any attempt is made to form a resonant packet. The absence of advanced waves can prevent energy transfer just as the absence of ground states can prevent emission of photons. The emission or absorption of energy by a PRAHM mode is therefore considered to be a statistical process similar to the statistical process of photon emission or absorption. It is again emphasised that advanced modes cannot initiate energy transfer which is controlled entirely by the retarded waves so that causality is never violated.

Consequently the suggestion is made that vector wavefunctions $\mathbf{\Psi_M}$ can be defined from the helically modulated retarded transverse fields alone. In a generic form one may write:

$$\mathbf{\Psi_M} = \Theta^{M+½}\,\mathbf{F_{TR}} \qquad \{\Theta^{M+½} = \exp[-(M+½)\,\sigma\omega(t - z/v_h)]\} \qquad 7.1$$

The different periods of helical rotation with different M ensure that the set $\{\mathbf{\Psi_M}\}$ are orthogonal when these vector functions are integrated over the cross section of the fibre and averaged over one period of the underlying mode of frequency $\omega$ (integration and averaging denoted by << >>). It is then possible to scale $\mathbf{F_{TR}}$ to define an orthonormal set $\{\mathbf{\Psi_M}\}$:

$$<<\mathbf{\Psi_M}^{\dagger}\mathbf{\Psi_N}>> = 0 \quad M \neq N\ ; \quad <<\mathbf{\Psi_M}^{\dagger}\mathbf{\Psi_M}>> = 1\ ; \quad (\dagger = \text{conjugation + transposition}) \qquad 7.2$$

As seen in equ. 3.9, the helical modulation has no effect on the classical energy just as the Schrödinger frequencies cancel when calculating the probability in quantum theory. It is now necessary to show how an equivalent energy equation to that of equation 6.12 can be made using 'wave-functions' of equ. 7.1.

Consider helical modulation and modal fields varying respectively as:

$$\Theta^{R} = \exp[R\,\sigma\omega(t - z/v_h)] \ : \ \mathbf{F_{TR}} = \mathbf{F_{TR0}}\cos(\omega t - kz + \phi). \qquad 7.3$$

The variables $(\sigma t)$ and $t$ are taken to be independent variables, analogous to real and imaginary parts of a complex variable. This independence is supported by the fact that $\omega$ and $\Omega$ are independent so that $\omega t$ and $\sigma\Omega t$ are independent. One can then formally write:

$$[\partial/\partial(\sigma t)]\ \mathbf{F_{TR}} = 0 \qquad 7.4$$

$$[\partial/\partial(\sigma t)]^{m}\ \Theta^{R}\ \mathbf{F_{TR}} = (\omega R)^{m}\ \Theta^{R}\ \mathbf{F_{TR}} \qquad 7.5$$

Extend this definition in equ.7.5 (with integer m) of the differential of an exponential function to include m = ½. Such derivatives are called pseudo-derivatives (Raymond 1991) and they have straightforward operational definitions provided one is dealing with exponential functions as one is here. It is then possible to define meaningful operators

$$A^{+} = \Theta^{½}\,[\partial/\partial(\sigma\omega t)]^{½}\,\Theta^{½} \qquad 7.6$$

$$A^{-} = \Theta^{-½}\,[\partial/\partial(\sigma\omega t)]^{½}\,\Theta^{-½} \qquad 7.7$$

Using equs. (7.1) and (7.4-7) it is possible to recover the well known relationships of QT:

$$A^{+}\boldsymbol{\Psi}_{\mathbf{M}} = (M+1)^{1/2}\, \boldsymbol{\Psi}_{\mathbf{M+1}} \;;\; A^{-}\boldsymbol{\Psi}_{\mathbf{M}} = M^{1/2}\, \boldsymbol{\Psi}_{\mathbf{M-1}} \;;\; A^{-}\boldsymbol{\Psi}_{\mathbf{0}} = 0 \qquad 7.8$$

$$A^{-}A^{+}\boldsymbol{\Psi}_{\mathbf{M}} = \Theta^{-1/2}\,[\partial/\partial(\sigma\omega t)]\,\Theta^{1/2}\,\Theta^{M+1/2}\,\mathbf{F}_{\mathbf{T}\mathrm{R}} = (M+1)\,\boldsymbol{\Psi}_{\mathbf{M}} \qquad 7.9$$

$$A^{+}A^{-}\boldsymbol{\Psi}_{\mathbf{M}} = \Theta^{1/2}\,[\partial/\partial(\sigma\omega t)]\,\Theta^{-1/2}\,\Theta^{M+1/2}\,\mathbf{F}_{\mathbf{T}\mathrm{R}} = M\,\boldsymbol{\Psi}_{\mathbf{M}} \qquad 7.10$$

$$A^{-}A^{+} - A^{+}A^{-} = 1 \qquad 7.11$$

$$\text{Energy} \propto \tfrac{1}{2}(A^{-}A^{+} + A^{+}A^{-}) = (M+\tfrac{1}{2}) \qquad 7.12$$

The operators $A^{-}$ and $A^{+}$ are similar to the operators in standard quantum theory (QT). With a scaled wavefunction these operators not only change the helical modulation of the wavefunction but provide an appropriate number operator $A^{+}A^{-}$ in equ. 7.10. This is proportional to $\partial/\partial(\sigma\omega t)$ which is analogous to the operation $i\partial/\partial(\omega t)$ determining the available energy in QT. Equ.7.12 is equivalent to the operation for the energy in QT and must match equ. 3.9. The counter helical rotation, regardless of the number state M, does not rotate the fields and so does not alter the matching of the vector fields at the boundaries of the waveguide. Because the standard QT formalism of the promotion and demotion operators is preserved so either Schrödinger and Heisenberg formalisms can be used here if required.

Annihilation in this model is straightforward. In equ. 7.8: $A^{-}\boldsymbol{\Psi}_{\mathbf{0}} = 0$ gives the usual mathematics of 'annihilation'. To look for a physical meaning consider a ground state of a PRAHM mode from:

$$\boldsymbol{\Phi}_{\mathbf{0}} = [\exp(\sigma\, \tfrac{1}{2}\omega\, \tau)\, \nabla\varphi_{\mathrm{R}} + \exp(-\sigma\, \tfrac{1}{2}\, \omega\, \tau)\, \exp(\sigma\, \phi)\, \nabla\varphi_{\mathrm{A}}\,]_{\,(-\tau_1 \le \tau \le \tau_2)} \qquad 7.13$$

Immediately after demoting the retarded field in the ground state one would find fields:

$$\boldsymbol{\Phi}_{\mathbf{0}} = \exp(-\sigma\, \tfrac{1}{2}\omega\, \tau)[\, \nabla\varphi_{\mathrm{R}} + \exp(\sigma\, \phi)\, \nabla\varphi_{\mathrm{A}}\,] \qquad 7.14$$

In equ. 7.14 there can be no trapping of the fields within any finite interval $-\tau_1 \le \tau \le \tau_2$ so that the energy in the packet must disperse or indeed be zero.

## 8. Discussion points and conclusions

### *8.1 Probability*

The path of a PRAHM mode is defined from the Maxwellian fields. The retarded fields, designated for example as $\boldsymbol{\Psi}$, match the boundary conditions and show the route that the PRAHM mode *has taken*. The advanced fields, designated for example as $\boldsymbol{\Psi}^*$, match future boundary conditions and determine all possible *future* route(s) for a PRAHM mode. Although acting as a pilot wave, the advanced $\boldsymbol{\Psi}^*$ is clearly different from the causal Bohmian pilot wave (Holland 1993). The PRAHM mode exists in the region of overlap of the retarded fields and advanced fields where the interaction energy is related to $\boldsymbol{\Psi}^{\dagger}\boldsymbol{\Psi}$. Thus $\boldsymbol{\Psi}^{\dagger}\boldsymbol{\Psi}$ can give a relative probability of finding the position of one PRAHM mode, allowing a conventional quantum probability interpretation. It is useful to observe that this work is also consistent with the quantum transactional theory of Cramer (1986) given some minor changes of interpretation: $\boldsymbol{\Psi}$ is the conventional (retarded) wave function while the conjugate $\boldsymbol{\Psi}^*$, travels with the same phase velocity as $\boldsymbol{\Psi}$, but is interpreted as an advanced wave. $\boldsymbol{\Psi}^{\dagger}\boldsymbol{\Psi}$ gives the relative probability of finding where, in space, retarded and advanced waves interact (overlap) and so $\boldsymbol{\Psi}^{\dagger}\boldsymbol{\Psi}$ gives the probability of the position of the particle (photon). With a one particle over a finite region, one normalises so that $\langle\boldsymbol{\Psi}^{\dagger}\boldsymbol{\Psi}\rangle = 1$.

*8.2 Uncertainty*

The fact that a localized wave-packet has been constructed from a mode with a definite angular frequency $\omega$ and wave-vector $\mathbf{k}$ does not invalidate the uncertainty principle. In the first instance the packet has a finite duration in time and space so that locating the exact position in time or space has an uncertainty. As the wave-packet of finite duration in space-time travels past an ideal measurement point, the Fourier analysis of the passing fields will necessarily yield the fundamental Fourier uncertainties in wave-vector and spatial position ($\Delta k\ \Delta x \sim 2\pi$). Similarly for any frequency and temporal observation period there will be the fundamental Fourier uncertainty ($\Delta\omega\ \Delta t \sim 2\pi$). It may be possible to extend the wave-packet for example by setting $\omega\tau_o = 2Q\pi$ with Q integer so that there are Q whole periods of the underlying mode. Although this will increase the certainty in $\omega$ and in $k$ it decreases the temporal and positional uncertainty. It is of course still possible, by superposition of different frequency states, to reduce the uncertainty in temporal or spatial position but this leads to greater uncertainty in $\omega$ or $k$. Uncertainty is fundamental to Fourier analysis and the formation of any wave-packet does not eliminate that uncertainty.

*8.3 Experimental support*

This present work uses the relative permeability $\varepsilon_r$ and permittivity $\mu_r$ as material parameters to describe the electron-material interaction. This classical approach, while similar to that used in some quantum theories of dielectrics (Huttner et al. 1991, Glauber and Lewenstein 1991) cannot deal with quantized loss or gain within the material as in more sophisticated theories (Matloob 2004 ). Nevertheless a new feature of this theory is the Schrödinger frequencies of (M+½)$\omega$ now have a clear meaning of giving helical modulations (local counter helical modulations to be exact) which have no effect on either the classical modal propagation or the polarisation. This explains how waves with quantum frequencies (M+½)$\omega$ can be associated with a classical wave of frequency $\omega$ travelling inside a dispersive waveguide without any change in that dispersion for the different values of M. Experiments [Ingham 2006] with single photons and classical fields in a fibre have suggested that, within experimental error, there are negligible differences between the single photon velocity and the classical energy velocity. This supports the concept that the photon velocity is substantially independent of the excitation number M. Such measurements of comparing single photon velocities with the classical group velocity have also been made in short lengths of bulk material [Steinberg 1992] and have similarly shown negligible differences within experimental limits.

It also seems possible that a theory can be constructed in free space that confines a photon-like wave-packet both transversely and longitudinally using, for example, some extension of the focus wave modes generalized through the use of TE and TM modes by Sezginer (1985) .

*8.4 Interference and Non-local features of localized wave-packets*

In the present theory, the energy is attached to the overlapping region of the counter rotating helical modulated retarded and advanced waves within the interval $-\tau_1 \le \tau \le \tau_2$ , ($\tau = t - z/v_h$ ; $\tau_o = \tau_1 + \tau_2$ ) . However the advanced waves $\mathbf{F}_{\mathrm{TA}}$ (valid in the region $\tau > -\tau_1$) are already present as background fields that satisfy Maxwell's equations and boundary conditions. These advanced waves are selected by the retarded waves $\mathbf{F}_{\mathrm{TR}}$ (valid in the region $\tau \le \tau_2$) in order to complete the resonant wave-packet. It is therefore these advanced modes that offer a way forward for rational explanations about the ability to predict future outcomes of interference or polarisation before the photon's energy arrives.

For the single PRAHM mode, advanced waves are solutions of Maxwell's equations but will have multiple paths, for example, when there are power splitters and combiners as in a Mach-Zehnder interferometer. The boundary conditions establish the phases and relative amplitudes of certain advanced waves along possible future paths of the PRAHM mode. Any attempt to measure where the PRAHM mode is located within the interferometer will destroy the established advanced waves and new advanced waves will have to be selected from the random background sea of advanced waves. These new advanced waves will not necessarily have the established phase relationships of the previous advanced waves. Consequently, even if the measurement has not removed (annihilated) the PRAHM mode, any future interference may be destroyed by the attempted measurement. All this is in accord with measurements of interference of a single photon within such an interferometer. It is therefore suggested that the advanced waves will be able to form an explanation of seminal work on delayed choice interference (Hellmuth et al.1987) and clever experiments on entangled photons (Shih and Alley 1988 ). Such experiments show that information about future phase or future polarisation of a photon can be conveyed in space-time ahead of the arrival of the photon's energy.

*8.5 Conclusions*

This new theory is believed to be the first model, based on the solid foundations of Maxwell's equations that offers such a good potential for resolving so many of the dilemmas posed by wave-particle duality within a relativistic context. It gives a reasonably intuitive account for a model photon that explains the significance of the Schrodinger frequencies (M+½)ω, the lack of additional dispersion and that these Schrödinger frequencies are unobserved because the classical energy and polarisation are unaffected by the counter helical modulations. It is believed that this is the first time that such a picture of *resonant* Schrödinger-like wavefunctions for a travelling photon has been put forward.

The PRAHM mode (**P**acket of **R**etarded and **A**dvanced **H**elically **M**odulated mode) is seen here as a quasi-classical model for a photon that has an energy of formation that is proportional to its helical modulation frequency of (M+½)ω consistent with Planck's law and causality. A heuristic model gives Planck's constant with the right order of magnitude. The wave-packet explains the localised properties of the photon while the advanced waves explain the non-local properties. The theory has survived its first theoretical tests of recovering the promotion and demotion operators as required for standard QT with Schrödinger and Heisenberg formulations. It has survived its first experimental test that suggests that there is little or no evidence that photons travel in a fibre at significantly different velocities in their excited (classical state M>>1) to their velocities as single photons (M=1): no additional dispersion. The theory appears capable of agreeing with Cramer's transactional interpretation of quantum theory. All this suggests that this new theory may be more than just another classical skirmish into quantum territory.

**Acknowledgements**

Thanks are given to Professors Bill Milne and Ian White for supporting the author's work in CAPE. Jonathan Ingham and Ruth Thompson have been particularly helpful in the experimental work. Joe Beales has significantly helped the author understand the meaning of the rotational modulation described here. Rodney Loudon and Seb Savory have given the author useful references.

**Appendix A. Maxwell's equations in matrix form for TE/TM fields**

Any spatial vector **F** and the gradient operator ∇ can be written in the form:

$$\mathbf{F} = (\mathbf{F_T}\ ,\ \mathbf{u}F_z)\quad \nabla = [\nabla_\mathbf{T},\ \mathbf{u}\partial_z] \qquad \text{A.1}$$

where **u** is the unit vector in the Oz direction and the subscript $_\mathbf{T}$ indicates the components transverse to Oz. Maxwell's divergence equs. may be written as:

$$\nabla_\mathbf{T}\ .\ c\mathbf{B_T} + \partial_z (cB_z) = 0\ ; \qquad \text{A.2}$$

$$\nabla_\mathbf{T}\ .\ c\mathbf{D_T} + \partial_z (cD_z) = c\rho\ ; \qquad \text{A.3}$$

Using the matrix notation of chapter 2, these equs. are re-arranged as

$$\nabla_\mathbf{T}{}^{tr}\ c\mathbf{B_T} + \partial_z (cB_z) = 0 \qquad \text{A.4}$$

$$\nabla_\mathbf{T}{}^{tr}\ \sigma^{tr}\sigma\ c\mathbf{D_T} + \partial_z (cD_z) = c\rho \qquad \text{A.5}$$

Setting $n^2$ = $\varepsilon_r$ and $\mu_r = 1$ with the approximation that all the energy is in the same value of $n^2$ over the whole fibre

$$(\sigma\nabla_\mathbf{T})^{tr}\ (\sigma\, n^2\ \mathbf{E_T}) + \partial_z (n^2\ E_z) = 0 \qquad \text{A.6}$$

Taking the transverse terms of Maxwell's two curl vector equations:

$$\partial_z (\mathbf{u} \times \mathbf{E_T}) - (\mathbf{u} \times \nabla_\mathbf{T})\ E_z + (1/c)\ \partial_t (c\mathbf{B_T}) = 0 \qquad \text{A.7}$$

$$\partial_z\, \mathbf{u} \times \mathbf{H_T} - (\mathbf{u} \times \nabla_\mathbf{T})\ H_z\ - (1/c)\ \partial_t (c\mathbf{D_T}) = -\ \mathbf{J_T} \qquad \text{A.8}$$

These respectively can be re-written in matrix form as

$$\partial_z (\sigma\ n^2\ \mathbf{E_T}) - \sigma\nabla_\mathbf{T}\ (n^2\ E_z) + (n^2/c)\ \partial_t (c\mathbf{B_T}) = 0 \qquad \text{A.9}$$

$$\partial_z (\sigma\ \mathbf{H_T}) - \sigma\nabla_\mathbf{T}\ (H_z) - (1/c)\ \partial_t (c\mathbf{D_T}) = -\ \mathbf{J_T} \qquad \text{A.10}$$

$$\partial_z (\sigma c\mathbf{B_T})\ - (n^2/c)\ \partial_t (\mathbf{E_T}\ ) - \sigma\nabla_\mathbf{T}\ (cB_z) = -c\mu_0\mu_r\ \mathbf{J_T} \qquad \text{A.11}$$

Now taking the Oz terms of Maxwell's curl equations (i.e. along direction **u**):

$$\mathbf{u}.(\nabla_\mathbf{T} \times \mathbf{E_T}) + (1/c)\ \partial_t (cB_z)\ = 0 \qquad \text{A.12}$$

$$\mathbf{u}.(\nabla_\mathbf{T} \times \mathbf{H_T}) - (1/c)\ \partial_t (\ cD_z)\ = -\ J_z \qquad \text{A.13}$$

These re-arrange into a matrix form:

$$-\nabla_\mathbf{T}{}^{tr}\ (\sigma\ \mathbf{E_T})\ + (1/c)\ \partial_t (cB_z)\ = 0 \qquad \text{A.14}$$

$$(\sigma\nabla_\mathbf{T}\ )^{tr}\ c\mu_0\mu_r\mathbf{H_T}\ -\ \partial_t (c\mu_0\mu_r\ \varepsilon_0\ \varepsilon_r E_z)\ = -\ c\mu_0\mu_r J_z \qquad \text{A.15}$$

which is re-written as:

$$(\sigma\nabla_\mathbf{T}\ )^{tr}\ (c\mathbf{B_T})\ - (1/c)\ \partial_t\ (n^2 E_z)\ = -\ c\mu_0\mu_r J_z \qquad \text{A.16}$$

Now re-grouping equations into those driven by $B_z$ and $(n^2 E_z)$ respectively:

$$\nabla_\mathbf{T}{}^{tr}\ c\mathbf{B_T} + \partial_z (cB_z) = 0 \qquad \text{A.17}$$

$$-\nabla_\mathbf{T}{}^{tr}\ (\sigma\ \mathbf{E_T})\ + (1/c)\ \partial_t (cB_z)\ = 0 \qquad \text{A.18}$$

$$\partial_z (\sigma c\mathbf{B_T})\ - (n^2/c)\ \partial_t (\mathbf{E_T}\ ) - \sigma\nabla_\mathbf{T}\ (cB_z) = -c\mu_0\mu_r\ \mathbf{J_T} \qquad \text{A.19}$$

$$(\sigma\nabla_\mathbf{T})^{tr}\ (\sigma\ n^2\mathbf{E_T}) + \partial_z (n^2 E_z) = 0 \qquad \text{A.20}$$

$$(\sigma\nabla_\mathbf{T}\ )^{tr}\ (c\mathbf{B_T})\ - (1/c)\ \partial_t\ (n^2 E_z)\ = -\ c\mu_0\mu_r J_z \qquad \text{A.21}$$

$$\partial_z \sigma\ n^2\mathbf{E_T} + (n^2/c)\ \partial_t (c\mathbf{B_T}) - \sigma\nabla_\mathbf{T}\ (n^2 E_z)\ = 0 \qquad \text{A.22}$$

These are Maxwell's equations in the matrix form required here.

**Appendix B. The advanced-retarded helical power theorem**

Consider retarded and advanced vector fields:

$$[\sigma\mathbf{E}_{\Theta\mathbf{TR}}, c\mathbf{B}_{\Theta\mathbf{TR}}, E_{zR}, cB_{zR}] = [\Theta\sigma\mathbf{E}_{\mathbf{TR}}, \Theta c\mathbf{B}_{\mathbf{TR}}, E_{zR}, cB_{zR}] \quad \text{B.1}$$

$$[\sigma\mathbf{E}_{\Phi\,\mathbf{TA}}, c\mathbf{B}_{\Phi\mathbf{TA}}, E_{zA}, cB_{zA}] = [\Phi\sigma\mathbf{E}_{\mathbf{TA}}, \Phi c\mathbf{B}_{\mathbf{TA}}, E_{zA}, cB_{zA}] \quad \text{B.2}$$

where $\Theta = \exp(\sigma\Omega\tau)$ $\Phi = \exp(-\sigma\Omega\tau) = \Theta^{-1}$ with $\tau = (t - z/v_g)$. *All* the underlying fields $\mathbf{E_T}$, $\mathbf{B_T}$, $E_z$, and $cB_z$ are assumed to have real variations $\cos(\omega t - k z + \phi)$.

There is an effective advanced current given from

$$\partial_z(\sigma c\mathbf{B}_{\mathbf{TA}}) - (n^2/c)\,\partial_t(\mathbf{E}_{\mathbf{TA}}) - \sigma\nabla_{\mathbf{T}}(cB_{zA}) = -c\mu_0\mu_r\mathbf{J}_{\mathbf{TeffA}} \quad \text{B.3}$$

$$(\sigma\nabla_{\mathbf{T}})^{tr}(c\mathbf{B}_{\mathbf{TA}}) - (1/c)\,\partial_t(n^2E_{zA}) = -c\mu_0\mu_r J_{zeffA} \quad \text{B.4}$$

Consider an increment $\Delta\tau$ of the transit time $\tau$. A positive $\Delta\tau$ for the retarded waves, can be matched by an equal but opposite $\Delta\tau$ for the advanced wave. Consequently we define a star operation * that reverses $\tau$ in the advanced waves so that $\Theta^{tr}\Phi^* = 1$ and allows one to calculate the work done against the effective advanced currents by the retarded fields. Using appendix A, the following equations can be formed:

$$\mathbf{E}_{\Theta\mathbf{TR}}{}^{tr}\partial_z(\sigma c\mathbf{B}_{\Phi\mathbf{TA}})^* - (n^2/c)\,\mathbf{E}_{\Theta\mathbf{TR}}{}^{tr}\partial_t(\mathbf{E}_{\Phi\mathbf{TA}}{}^*) - \mathbf{E}_{\Theta\mathbf{TR}}{}^{tr}\sigma\Phi^*\nabla_{\mathbf{T}}(cB_{zA}{}^*)$$
$$= -c\mu_0\mu_r\mathbf{E}_{\Theta\mathbf{TR}}{}^{tr}\Phi^*\mathbf{J}_{\mathbf{TeffA}}{}^* \quad \text{B.5}$$

$$(\partial_z\sigma\,\mathbf{E}_{\Theta\mathbf{TR}}{}^{tr})c\mathbf{B}_{\Phi\mathbf{TA}}{}^* + (1/c)\,\partial_t(c\mathbf{B}_{\Theta\mathbf{TR}}{}^{tr})\,c\mathbf{B}_{\Phi\mathbf{TA}}{}^* - (\sigma\,\Theta\nabla_{\mathbf{T}})^{tr}(E_{zR})\,c\mathbf{B}_{\Phi\mathbf{TA}}{}^* = 0 \quad \text{B.6}$$

$$-cB_{zA}{}^*\nabla_{\mathbf{T}}{}^{tr}(\sigma\,\mathbf{E_T}) + cB_{zA}{}^*(1/c)\,\partial_t(cB_z) = 0 \quad \text{B.7}$$

$$(E_{zR})(\sigma\nabla_{\mathbf{T}})^{tr}(c\mathbf{B}_{\mathbf{TA}}{}^*) - (E_{zR})(1/c)\,\partial_t(n^2E_{zA}{}^*) = -(E_{zR})c\mu_0\mu_r J_{zA}{}^* \quad \text{B.8}$$

The $\nabla_{\mathbf{T}}$ terms are changed by use of the identies:

$$-\mathbf{E}_{\Theta\mathbf{TR}}{}^{tr}\sigma\Phi^*[\nabla_{\mathbf{T}}(cB_{zA}{}^*)] = \nabla_{\mathbf{T}}[(\sigma\mathbf{E}_{\mathbf{TR}})^{tr}\; cB_{zA}{}^*] - cB_{zA}{}^*\nabla_{\mathbf{T}}{}^{tr}(\sigma\mathbf{E}_{\mathbf{TR}}) \quad \text{B.9}$$

$$-[(\sigma\nabla_{\mathbf{T}})^{tr}(E_{zR})]\,c\mathbf{B}_{\mathbf{TA}}{}^* = -(\sigma\nabla_{\mathbf{T}})^{tr}[(E_{zR})\,c\mathbf{B}_{\mathbf{TA}}{}^*] + (E_{zR})[(\sigma\nabla_{\mathbf{T}})^{tr}c\mathbf{B}_{\mathbf{TA}}{}^*] \quad \text{B.10}$$

Then integration of the cross section (denoted by $\langle\,\rangle$) eliminates the following terms

$$\langle\nabla_{\mathbf{T}}[(\sigma\mathbf{E}_{\mathbf{TR}})^{tr}\; cB_{zA}{}^*]\rangle = 0 \;\; ; \;\; \langle(\sigma\nabla_{\mathbf{T}})^{tr}[(E_{zR})\,c\mathbf{B}_{\mathbf{TA}}{}^*]\rangle = 0 \quad \text{B.11}$$

Averaging over the cross section then permits one to write:

$$\langle\,\partial_z[\mathbf{E}_{\mathbf{TR}}{}^{tr}(\sigma c\mathbf{B}_{\mathbf{TA}})^*] - \Omega\,(n^2/c)\,\mathbf{E}_{\mathbf{TR}}{}^{tr}\sigma\,\mathbf{E}_{\mathbf{TA}}{}^* + \Omega(1/c)(c\mathbf{B}_{\mathbf{TR}}{}^{tr}\sigma\,c\mathbf{B}_{\mathbf{TA}}{}^*)$$
$$-(1/c)\,[\partial_t(c\mathbf{B}_{\mathbf{TR}}{}^{tr})]\,c\mathbf{B}_{\mathbf{TA}}{}^*$$
$$-(n^2/c)(\mathbf{E}_{\mathbf{TR}}{}^{tr}\partial_t\mathbf{E}_{\mathbf{TA}}{}^*) - cB_{zA}{}^*(1/c)\,\partial_t(cB_z) - (E_{zR})(1/c)\,\partial_t(n^2E_{zA}{}^*)\rangle$$
$$= \langle -c\mu_0\mu_r\mathbf{E}_{\Theta\mathbf{TR}}{}^{tr}\Phi^*\mathbf{J}_{\mathbf{TA}}{}^* - (E_{zR})c\mu_0\mu_r J_{zA}{}^*\rangle \quad \text{B.12}$$

Re-arranging and integrating over one whole temporal period as well as the cross section (denoted by $\langle\langle\;\rangle\rangle$) yields a helical power flow theorem analogous to the complex power flow theorem (equ. 6.3)

$$\langle\langle -\partial_z[\mathbf{E}_{\mathbf{TR}}{}^{tr}\sigma\;\mathbf{H}_{\mathbf{TA}}{}^*] + \Omega\,[\mathbf{D}_{\mathbf{TR}}{}^{tr}\sigma\mathbf{E}_{\mathbf{TA}}{}^* - \mathbf{H}_{\mathbf{TR}}{}^{tr}\sigma\mathbf{B}_{\mathbf{TA}}{}^*]\rangle\rangle$$
$$= \langle\langle\,\mathbf{E}_{\mathbf{TR}}{}^{tr}\mathbf{J}_{\mathbf{TA}}{}^* + (E_{zR})\,J_{zA}{}^*\rangle\rangle \quad \text{B.13}$$

Note that the advanced and retarded fields are not necessarily identical to each other so that this theorem requires interpretation as given in section 6.